\documentclass{article}

\usepackage{graphicx}

 \newcommand{\ctg}{\mathop{\rm ctg}\nolimits}

\begin{document}

\title{The cyclotron spectrum of anisotropic ultrarelativistic electrons:
interpretation of X-ray pulsar spectra.}

\author{A.N. Baushev \thanks{Space Research Institute, Russian Academy of Sciences, Profsoyuznaya str. 84/32, Moscow, 117810 Russia\endgraf
Bogoliubov Laboratory of Theoretical Physics, Joint Institute for Nuclear Research;
141980 Dubna, Moscow Region, Russia\endgraf
Email: abaushev@mx.iki.rssi.ru}}
\date{}

\maketitle

\begin{abstract}
The spectrum of cyclotron radiation produced by electrons with a strongly anisotropic velocity
distribution is calculated taking into account higher harmonics. The motion of the electrons is assumed to
be ultrarelativistic along the magnetic field and nonrelativistic across the field. One characteristic feature of
the resulting spectrum is that harmonics of various orders are not equally spaced. The physical properties
and observed spectra of four X-ray pulsars displaying higher cyclotron harmonics are analyzed. The
cyclotron features in the spectra of all four pulsars can be interpreted only as emission lines. Moreover,
the observed harmonics are not equidistant, and display certain other properties characteristic of emission
by strongly anisotropic ultrarelativistic electrons. In addition, there are indirect theoretical arguments
that the electrons giving rise to cyclotron features in the spectra of X-ray pulsars are ultrarelativistic
and characterized by strongly anisotropic distributions. As a result, estimates of the magnetic fields of
X-ray pulsars (which are usually derived from the energies of cyclotron lines) and certain other physical
parameters require substantial revision.
\end{abstract}

\section{INTRODUCTION}

A cyclotron line in the spectrum of an X-ray pulsar was first identified for the object Her X-1 in 1977 \cite{g}.
Reports of observations of higher cyclotron harmonics in the spectra of several X-ray pulsars in binary systems
(X0115+63 \cite{a}; Vela X-1 \cite{c},\cite{c1}; 4U1907+09 \cite{c2},\cite{c3}; A0535+26 \cite{d})\footnote{Only references
to the first publications for each source are
given here.} were published later. Two cyclotron harmonics were also observed in the spectrum of 1E 2259+586 \cite{d11}; however,
the nature of this object remains unknown, and will not be discussed here. At least three harmonics have been detected
in the spectrum of X0115+63 \cite{d1, b}. The presence of higher cyclotron harmonics seems to be established most firmly for this source.
Unfortunately, the situation for the other objects is less clear, since only weak second harmonics have
been detected in their spectra. Nevertheless, even these first observations are very interesting, since they
can be used to address many theoretical problems connected with the radiation of X-ray pulsars. The aim of the present paper is to consider the
physical conditions for the formation of higher cyclotron harmonics in pulsar spectra. We find that a number of problems related to the
structure of the emitting region of these sources that have remained poorly understood can be easily solved. Here is a list of
such problems, which will be discussed in more detail below.

First, it remains unknown if the cyclotron features in the spectra of many X-ray pulsars are absorption
or emission lines. We will argue below that the four sources indicated above display emission lines (of
course, if higher-order harmonics are indeed present in their spectra).

Second, the velocity distribution of the electrons emitting the cyclotron lines is not entirely clear.
We can reasonably assume that the velocity of their motion across the magnetic field is fairly low, i.e.,
weakly relativistic (or even nonrelativistic). Otherwise, the cyclotron line would have the characteristic
synchrotron shape, with a large number of harmonics forming a quasi-continuum. In fact, we can usually
see only one (the fundamental) harmonic. A few exceptions are listed above, but these sources also
possess moderate transverse velocities. The electron velocities along the magnetic field are unclear. It
is usually assumed that these velocity are small, considerably less than $c$.
Nevertheless, it is possible that the electron-velocity distributions are strongly
anisotropic, so that the electron velocities are weakly relativistic across the magnetic field and ultrarelativistic
along the field. As was shown in the analysis of the spectrum of the X-ray pulsar Her X-1 in \cite{i},
there are strong arguments suggesting that the
electrons forming the cyclotron line are ultrarelativistic,
and that their velocities are characterized by
very strong anisotropy\footnote{Here and below, a strongly anisotropic distribution will be
defined as a velocity distribution for electrons whose motion
along the magnetic field is ultrarelativistic and across the
field is nearly nonrelativistic. This subject will be discussed
in more detail below.}. Unfortunately, if only one cyclotron harmonic is observed (as for Her X-1), full
verification of this possibility purely through spectral analysis is not possible, so that some additional
considerations (the dependence of the spectrum on the phase of the pulsar, and so on) must be used. The situation is completely different if several
harmonics are observed. As will be shown below, cyclotron radiation by ultrarelativistic, anisotropic electrons displays
very specific features. In particular, the harmonics of various orders are not equally spaced. Therefore, by analyzing the spectrum, we can determine
unambiguously if the motion of the emitting electrons along the field is ultrarelativistic or not.

We will use the following model for the emitting region ("hot spot") of an X-ray pulsar, put forward in \cite{i}.
When the accretion flow approaches the surface of the neutron star, a shock wave is formed at some
distance from this surface, and a turbulent region with temperature $T_e$ forms below the shock. The heated
surface of the neutron star, characterized by the temperature $T_s$, is located below the turbulent region. The
ultrarelativistic electrons emitting the cyclotron line are produced at the shock front; their optical depth is assumed to be negligible.

\section{CALCULATION OF THE RADIATION SPECTRUM IN A COMOVING COORDINATE FRAME}

Let us calculate the magnetobremsstrahlung radiation from electrons located in a constant magnetic field and possessing a strongly anisotropic
distribution. We assume that the electron velocities along the magnetic-field direction (denoted $V$) are all equal,
and that the motion of the electrons along the magnetic field is ultrarelativistic. (The situation for the transverse distribution will be
specified below.) Let us use a reference frame moving along the magnetic field with a constant velocity $V$. Although this coordinate
system is not fully associated with the electrons, we will call it the comoving frame, while the initial fixed
reference frame will be called the laboratory frame. The electrons have only a transverse velocity component $\upsilon$ in the comoving frame.
It is assumed in our model that $\upsilon \ll c$, and the velocity distribution is of
the form\footnote{We emphasize that, here and in (\ref{j6}), $\upsilon$ is the total velocity in
the comoving frame, not the velocity component perpendicular to the magnetic field in the laboratory frame.}:
\begin{equation}
\label{s0}
dn = N exp\left(-\frac{m_e \upsilon^2}{2 T}\right)\,
d~\left(-\frac{m_e \upsilon^2}{2 T}\right)
\end{equation}
In other words, this is a nonrelativistic, two-dimensional,
Maxwellian distribution with temperature
$T \ll m_e c^2$. Let us first find the radiation field in the
comoving frame. Further, we will be able to derive
the distribution of radiation in the laboratory frame by
performing a Lorentz transformation.

According to \cite{f}, the emission of each harmonic in the nonrelativistic case occurs at the fixed frequency
\begin{equation}
\label{f5}
\omega_n = n \omega_H , \mbox{where} \omega_H = \frac{e H}{m_e c} \mbox{ is the Larmor frequency}
\end{equation}
and the intensity of the radiation at the $n$th harmonic emitted by a particle moving with
velocity $\upsilon \ll c$ across the magnetic field $H$, is given by the formula:
\begin{equation}
\label{s1}
F'_n=\frac{n^2 e^4 H^2}{2 \pi m^2_e c^2} \left(1-\frac{\upsilon^2}{c^2}\right) \left[ \ctg^2\theta'
J^2_n\left( n \frac{\upsilon}{c} \sin\theta' \right) + \frac{\upsilon^2}{c^2} \dot J^2_n\left( n \frac{\upsilon}{c}
\sin\theta'\right)\right]
\end{equation}
where $\theta'$ is the angle between the magnetic-field vector and the direction toward the observer, $J$ and $\dot J$ are
the corresponding Bessel function and its derivative, and a prime denotes quantities measured in the comoving frame.
Here and below, the intensity is defined as the energy emitted by the system into unit solid angle per unit time.

Bessel functions can be expanded into standard Taylor series about zero:
\begin{eqnarray}
\label{s2}
J_n=\sum_{\nu=0}^\infty \frac{(-1)^\nu}{\nu! (n+\nu)!} \left( \frac{x}{2}
\right)^{n+2\nu} ,\qquad
\dot J_n=\sum_{\nu=0}^\infty \frac{(-1)^\nu (n+2\nu)}{2 (\nu!) (n+\nu)!}
\left( \frac{x}{2} \right)^{n+2\nu-1}
\end{eqnarray}

Thus, if the quantity
\begin{equation}
\label{s3}
n \frac{\upsilon}{c} \cos\theta'
\end{equation}
is not very big, we can take to account only the first term of the expansion. After some transformations we obtain:
\begin{eqnarray}
\label{s4}
\left[ \ctg^2\theta' J^2_n\left( n \frac{\upsilon}{c} \sin\theta' \right) +
\frac{\upsilon^2}{c^2} \dot J^2_n \left( n \frac{\upsilon}{c} \sin\theta'
\right)\right] \simeq \nonumber\\
\simeq\frac{1}{n!} \left( \frac{n \upsilon}{2 c} \sin\theta'
\right)^{2 n} \left(\frac{2-\sin^2\theta'}{\sin^2\theta'} \right)
\end{eqnarray}
Moreover, in this approximation $ \left(1-\frac{\upsilon^2}{c^2}\right)
\simeq 1$, and from (\ref{s1}) we get:
\begin{equation}
\label{s5}
F'_n=\frac{3 n^2 \sigma_T c H^2}{16 \pi^2} \frac{1}{n!}
\left( \frac{n \upsilon}{2 c} \sin\theta \right)^{2 n}
\left(\frac{2-\sin^2\theta'}{\sin^2\theta'} \right)
\end{equation}
where the Thomson scattering cross section
$\displaystyle{\sigma_T=\frac{8\pi}{3}\left(\frac{e^2}{m_e c^2}\right)^2}$
has been introduced. To obtain the total intensity of the radiation emitted by $N$ particles distributed
in accordance with (\ref{s0}), we must integrate (\ref{s5}) over the distribution:
\begin{equation}
\label{s6}
I'_n=\int\limits_0^\infty F'_n\,d\frac{m_e\upsilon^2}{2 T}
\end{equation}
After dividing this integral by $N$, we obtain the average intensity of the radiation emitted by one particle:
\begin{equation}
\label{d1}
I'_n=\frac{3\sigma_T c H^2}{16 \pi^2}\frac{n^{2n+2}}{n!}
(\sin\theta')^{2n} \frac{(2-\sin^2\theta')}{\sin^2\theta'}
\left(\frac{T}{2 m_e c^2}\right)^n
\end{equation}
We can see that $\displaystyle{\lim_{n\to\infty}I'_n=\infty \mbox{, when } \theta' \ne
\frac{\pi}{2}}$. Obviously, this is related to the limited applicability of the
asymptotic expansions of the Bessel function used in (\ref{s4}). In fact, expression
(\ref{d1}) represents a zeroth approximation to the intensity. Taking into account
the following terms of the expansion, the condition for
the applicability of (\ref{d1}) can be formulated as:
\begin{equation}
\label{d7}
\frac{n (n+2)}{(n+1)}\frac{T}{m_e c^2}\ll 0.3
\end{equation}
Formula (\ref{d1}) yields substantially overestimated values
for large transverse temperatures, and is not applicable
at all when
\begin{equation}
\label{d8}
\frac{n (n+2)}{(n+1)}\frac{T}{m_e c^2}\approx 1
\end{equation}
As is noted above, the radiation intensities in the laboratory and comoving frames are related by an ordinary
Lorentz transformation. The comoving frame moves with respect to the laboratory frame with the
velocity $V$ . Let us introduce the notation\footnote{As in the previous formulas, primed quantities refer to the
comoving frame, and unprimed quantities refer to the laboratory
frame or are frame independent.}:
\begin{equation}
\label{d9}
\gamma=\frac{1}{\sqrt{1-\frac{V^2}{c^2}}}\qquad
\mu=\frac{1}{1-\frac{V}{c}\cos\theta}\qquad
\hat\mu=\frac{1}{1+\frac{V}{c}\cos\theta'}
\end{equation}
Using the results of \cite{f}, we can easily derive the relations:
\begin{equation}
\label{d10}
\omega=\omega' \frac{\sqrt{1-\frac{V^2}{c^2}}}{1-\frac{V}{c}\cos\theta}=
\omega' \frac{\mu}{\gamma}
\end{equation}
\begin{equation}
\label{d11}
\cos\theta=\frac{\cos\theta'+\frac{V}{c}}{1+\frac{V}{c}\cos\theta'}=\hat
\mu(\cos\theta'+\frac{V}{c})
\end{equation}
We also emphasize the important relation
\begin{equation}
\label{d12}
\mu\hat\mu=\gamma^2
\end{equation}
Using (\ref{d9} - \ref{d12}), we can easily obtain the relations:
\begin{equation}
\label{d13}
\sin\theta'=\frac{\gamma}{\hat \mu}\quad \sin\theta=\frac{\mu}{\gamma}\quad
\sin\theta do'= \frac{\mu^2}{\gamma^2} do,
\end{equation}
where $do$ and $do'$ are elements of solid angle in the laboratory and comoving frames.

Next, according to \cite{f}, the transformation of the intensity from the comoving to the laboratory frame takes the form:
\begin{equation}
\label{e2}
I=\frac{\mu^3}{\gamma^2} I'
\end{equation}
Radiation at some specific $n$th harmonic in the comoving frame is monochromatic with frequency $n\omega_H$
In the laboratory frame, the frequency of the radiation emitted by one particle is an single-valued function of
the observation angle $\theta$ è (and, therefore, of $\theta'$). Let us find this function in explicit form. According to (\ref{d10}),
$\displaystyle{\omega=\frac{\mu}{\gamma}\omega'}$. Substituting $n\omega_H$ instead of $\theta'$ into this and
using (\ref{d12}), we obtain:
\begin{equation}
\label{e3}
\omega=\frac{\mu}{\gamma} n \omega_H=\frac{\gamma}{\hat\mu}n\omega_H=(1+\frac{V}{c}\cos\theta')\gamma n\omega_H
\end{equation}
If $V\simeq c$ and the angles $\theta'$ are not very close to $\pi$, we
have\footnote{The condition for smallness of the angle, when
(\ref{e4}), is applicable, can be written:
$$
\frac{(1+\cos\theta')-(1+\frac{V}{c}\cos\theta')}{(1+\frac{V}{c}\cos\theta')}
\ll 1
$$
If $\frac{V}{c}\simeq 1$ this condition is not valid only for angles $\theta'$
such that $\displaystyle{\pi-\theta'\le\frac{1}{\gamma}}$. We can see that the solid angle corresponding to
these plane angles is extremely small $\left(\displaystyle{\sim\frac{\pi}{\gamma^2}}\right)$.
Since the intensities of all harmonics in the comoving frame have no features when $\theta'\simeq\pi$,
the relative fraction of radiation corresponding to this solid angle is obviously also very small
$ \left(\sim\frac{1}{4 \gamma^2} \right) $, and can be neglected.}:
\begin{equation}
\label{e4}
(1+\frac{V}{c}\cos\theta')\approx 1+\cos\theta'
\end{equation}
We obtain in the above approximation
$$
\omega=(1+\cos\theta')\gamma n \omega_H
$$
$$
\cos\theta'=\frac{\omega}{\gamma n \omega_H}-1
$$
\begin{equation}
\label{e5}
\sin\theta'= 1-\left(\frac{\omega}{\gamma n \omega_H}-1\right)^2=
2\frac{\omega}{\gamma n \omega_H}-\left(\frac{\omega}{\gamma n
\omega_H}\right)^2
\end{equation}
We emphasize that (\ref{e5}) is valid only when $\displaystyle{\tilde \omega \in [0; 2n\gamma\omega_H]}$
(since $\theta' \in [0; \pi]$, so that $\sin \theta' \in [0; 1]$).

Since the frequency of the radiation is a single valued function of the angle, we can use this frequency,
instead of the angle, as the variable specifying the direction. Substituting expression (\ref{d1}) for the
intensity $I'_n$ into (\ref{e2}), replacing $\sin\theta'$ in accordance with (\ref{e5}), and using the relation (\ref{d10}),
we finally obtain:
\begin{eqnarray}
\label{f2}
I_n=\frac{3 \gamma \sigma_T c H^2}{16 \pi^2} \frac{n^{2n+2}}{n!}
(\frac{\omega}{n \omega_H})^3 \left(2\frac{\omega}{\gamma n \omega_H}-
\left(\frac{\omega}{\gamma n \omega_H}\right)^2\right)^{n-1}\times\nonumber\\
\times\left(1+\left(\frac{\omega}{\gamma n \omega_H}-1\right)^2\right)
\left(\frac{T}{2 m_e c^2}\right)^n
\end{eqnarray}
This formula represents the intensity of the radiation emitted by one particle at the $n$th harmonic,
where the directional dependence of the intensity is expressed in terms of the frequency emitted in that direction.

\section{CALCULATION OF THE OBSERVED CYCLOTRON SPECTRUM}
Let us now find the radiation spectrum detected by an observer an infinite distance from the star.
It is known (see, for example, \cite{v})that magnetobremsstrahlung
radiation by ultrarelativistic particles
is highly directional, and is nearly completely concentrated
in a narrow cone (with an opening angle of about $\simeq
\frac{1}{\gamma}$) along the direction of the particle's
velocity (in the case under consideration, along the
magnetic field). Therefore, the radiation detected by
an observer at an infinite distance at any particular
time is produced by a very small part of the emitting
region, where the magnetic field is directed precisely
toward the observer. The linear size of this region is of
the order of $R \frac{1}{\gamma}$, where $R$ is the star's radius.

Let us assume that the electrons are uniformly
distributed over this part of the emitting region with
surface density $\rho$, that their distribution function has
the form (\ref{s0}), and that the magnetic field is constant
and perpendicular to the stellar surface. Let
this surface be separated into bands of width $R d\varphi$,
which are symmetric with respect to the line from the
neutron-star center to the observer. Here, $\varphi$ is the
angle between rays drawn from the stellar center to
the observer and a given point on the stellar surface
(i.e., the latitude of this point). Let us consider one of
these bands. Its area is
$ 2 \pi R^2 \sin\varphi d\varphi $, and the number
of emitting electrons in this band is
\begin{equation}
\label{e61}
\rho \cdot 2 \pi R^2 \sin\varphi d\varphi
\end{equation}
Since we have assumed above that the magnetic field
is everywhere perpendicular to the stellar surface, the
angle between the normal to this surface and the
direction toward an observer is equal to the angle between
the magnetic field and the direction toward the
observer. As follows from geometrical considerations,:
\begin{equation}
\label{e7}
\varphi=\theta \quad \mbox{and}\quad d\varphi=d\theta
\end{equation}
The frequency of the radiation emitted by the band is
given by (\ref{e3}), where $\theta$ can be replaced by $\varphi$:
\begin{equation}
\label{e8}
\omega=\frac{\mu}{\gamma} n \omega_H=\frac{n\omega_H}{\gamma
(1-\frac{V}{c}\cos\varphi)}
\end{equation}
The angle $\varphi$ is not absolutely constant within the
band, and changes by an amount $d\varphi$. As a result, the
radiation emitted by the band is not monochromatic, and covers some frequency interval $d\omega$.
Let us calculate
the width of this interval. Differentiating (\ref{e8}) with
respect to $\varphi$, we obtain for  $\frac{V}{c}\sim 1$:
\begin{equation}
\label{e10a}
d\omega=n \omega_H\frac{\mu^2}{\gamma} \sin{\varphi}d\varphi
\end{equation}
Therefore, the band contributes to the total spectrum
of the emitting region in the frequency interval $[\omega-d\omega ; \omega]$, where
$\omega$ and $d\omega$ are given by (\ref{e8}) and (\ref{e10a}). Moreover, it can easily
be deduced that the corresponding interval for the total spectrum is produced
only by electrons in this band.
As follows from (\ref{e10a})
\begin{equation}
\label{e11}
\sin{\varphi}d\varphi=d\omega=\frac{\mu^2}{\gamma}\frac{d\omega}{n \omega_H}
\end{equation}

Substituting (\ref{e11}) into (\ref{e61}), we obtain for the number
of particles $dq$ emitting in the band:
\begin{equation}
\label{e12}
dq=2 \pi \rho R^2 \frac{\gamma}{\mu^2}\frac{d\omega}{n \omega_H}
\end{equation}
Let do be the solid angle subtended by the observer
at the neutron-star surface. Then, the total energy
received by the observer from the band under consideration
during a time $dt$ will be
$$
dE = I dq dt do
$$
where $I$ is the intensity emitted by one particle. Substituting
(\ref{e12}) for $dq$, using (\ref{d10}), and dividing both
sides of the equality by $d\omega do dt$, we obtain an expression
for the spectral energy density from the star $P_n$, i.e., for the amount of energy emitted by the star per
unit frequency interval into unit solid angle per unit time:
\begin{equation}
\label{f1}
P_n = I \frac{2 \pi \rho R^2}{\gamma} \frac{n \omega_H}{\omega^2}
\end{equation}
Dividing this expression by $\hbar \omega$, we make the transformation
from the spectral energy density $P_n$ to the
spectral particle-flux density $Q_n$. Let us introduce the
dimensionless frequency
\begin{equation}
\label{j1}
\tilde \omega = \frac{\omega}{\gamma \omega_H}
\end{equation}
Substituting formula (\ref{f2}) for the cyclotron intensity
emitted by one particle into (\ref{f1}) and using the relation
$$
\frac{3 c \sigma_T  H^2}{8 \pi \hbar {\omega_H}^2} = \frac{e^2}{\hbar c} =
\lambda \simeq \frac{1}{137}
$$
(where $\lambda$ is the fine-structure constant), we finally
obtain:
\begin{equation}
\label{f4}
Q_n= \lambda \rho R^2 \frac{n^{2n}}{n!}
\left(2\frac{\tilde \omega}{n}-
\left(\frac{\tilde \omega}{n}\right)^2\right)^{n-1}
\left(1+\left(\frac{\tilde \omega}{n}-1\right)^2\right)
\left(\frac{T}{2 m_e c^2}\right)^n
\end{equation}
This formula determines the number of photons of the
$n$th harmonic emitted at some time by the star per
unit frequency interval into unit solid angle per unit
time (i.e., the instantaneous observed spectrum of the
pulsar). If the physical conditions are the same over
the entire surface of the spot, this spectrum coincides
with the time-averaged spectrum of the pulsar, with
a coefficient corresponding to the fraction of time the
hot spot is observed. Since the transformation (\ref{e5}) is
valid for $\tilde \omega \in [0; 2n]$, formula (\ref{f4}) is valid when
\begin{equation}
\label{j4}
0 \le \tilde \omega \le 2 n
\end{equation}
Physically, this inequality reflects the fact that, according to (\ref{d10}),
the photon frequency cannot increase by more than the factor
$2 \gamma$ in the transformation from the comoving to the laboratory frame, and all the photons
have the frequency $\omega_n = n \omega_H$ in the comoving
frame. Therefore, there are no photons with frequencies
exceeding $2 \gamma n \omega_H$ in the laboratory harmonic
spectrum. Precisely this restriction is responsible for
the sharp cut-off in the spectrum of the first harmonic.

We have neglected the gravitational redshift when
deriving (\ref{f4}) and in subsequent relations. When this
redshift is fairly small (the stellar radius $R$ is far from
the gravitational radius $r_g$), it can easily be taken into
account by substituting the quantity
\begin{equation}
\label{j3}
\Omega = \frac{\omega}{\gamma \omega_H} \sqrt{1-\frac{r_g}{R}}
\end{equation}
in place of the frequency given by (\ref{j1}). Nevertheless,
we shall primarily use the previously derived formula
(\ref{f4}).

Combining harmonics of several orders (beginning with the first) and using the constraint
(\ref{j4}), we obtain the resulting cyclotron spectrum. Here
and below, only the first four cyclotron harmonics
will be used to construct the spectra, while lines of
higher orders will be neglected. Let us assume that
the distance to the pulsar is $r=3.5$~{kpc}, the radius
of the neutron star is $R=10~{km}$ (these correspond
approximately to the parameters of X0115+63), and
that the surface density of the radiating particles is $\sim 10^{16}~{cm^{-2}}$.
The value of $\hbar \gamma \omega_H$ will be taken to
be $10~{keV}$, which is also in reasonable agreement
with the energy of the first harmonic observed in the
spectrum of X0115+63 at $\sim 20~{keV}$.

\begin{figure}
\vspace*{-1cm}
\resizebox{0.75\hsize}{!}{\includegraphics[angle=270]{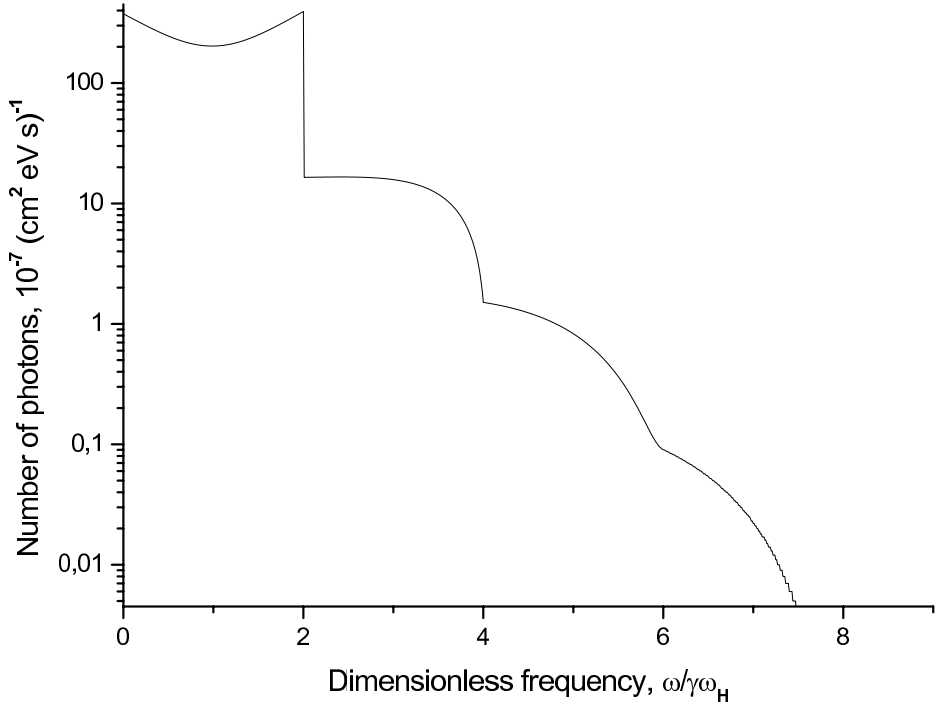}}
\caption{Spectrum of cyclotron radiation emitted by an X-ray
pulsar containing four harmonics, calculated using
(\ref{f4}). The transverse temperature of the emitting electrons
is $T = 20$~{keV}. The dimensionless frequency is plotted on
a linear scale along the horizontal axis, and the number of
particles on a logarithmic scale along the vertical axis.}
\label{fig1}
\end{figure}

The energy distribution of the photons that should
be detected by an observer on the Earth for a transverse
temperature of the emitting electrons $T = 20$~{keV}
is presented in Fig.~\ref{fig1}. We can see that these
harmonics are very broad, and overlap each other. In
addition, they are clearly not equally spaced; in particular,
the maximum of the second harmonic nearly
coincides with the maximum of the first harmonic.
The first harmonic possesses a second maximum
at $\omega \simeq 0$, which is somewhat weaker than the main
maximum, since higher harmonics are also added to
the main maximum.

\begin{figure}
\vspace*{-1cm}
\resizebox{0.75\hsize}{!}{\includegraphics[angle=270]{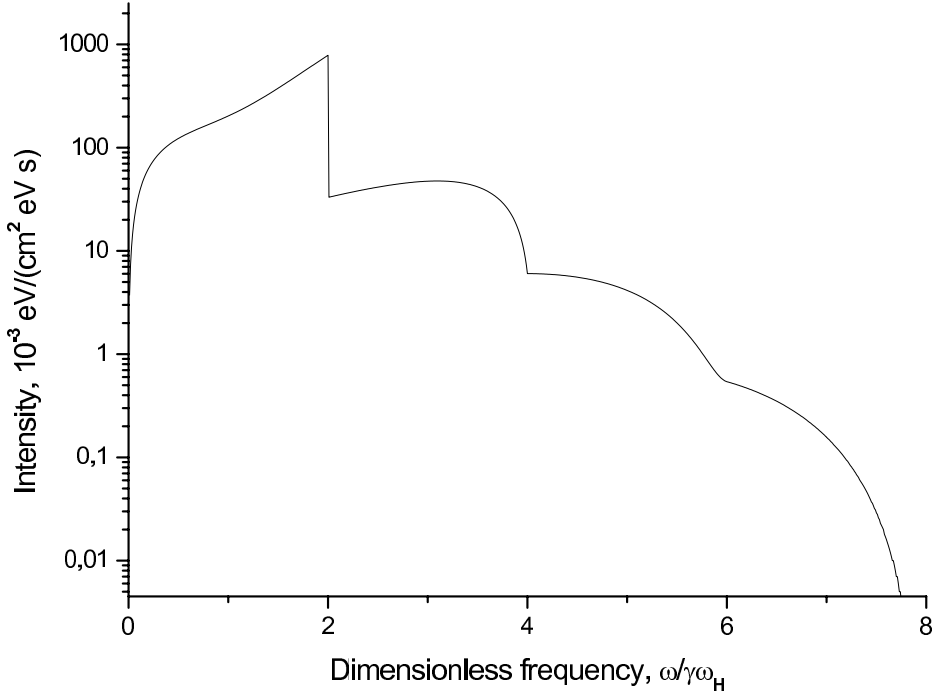}}
\caption{Energy spectrum of cyclotron radiation by an
X-ray pulsar containing four harmonics. The transverse temperature
of the emitting electrons
is $T = 20$~{keV}. The dimensionless frequency is plotted on
a linear scale along the horizontal axis, and the number of
particles on a logarithmic scale along the vertical axis.}
\label{fig2}
\end{figure}

It is usual to depict the frequency dependence of
the total photon energy, rather than of the number
of photons. Figure~\ref{fig2} presents this dependence for
the same transverse temperature $T = 20$~{keV}. The
formula describing this spectrum can be obtained
by multiplying (\ref{f4}) by $\hbar \omega$. As a result, the maxima
of the harmonics will be shifted; in particular, the
maximum of the second harmonic will not coincide
with the maximum of the first harmonic, and will
appear distinct. In addition, the maximum at $\omega \simeq 0$
disappears. However, as before, the spectral features
are not equidistant, and the harmonics are very broad
and overlap each other.

These effects have a simple qualitative explanation.
Let us consider the emission of photons in the
comoving frame. First, we must answer the question
of which photons will reach the distant observer. It
is obvious that these will be photons movingaway
from the star in the laboratory frame; i.e., those for
which $\theta\le\frac{\pi}{2}$. According to (\ref{d10}), this corresponds to
the angles $0\le\theta'\le\pi-\frac{1}{\gamma}$ ) in the comoving
frame\footnote{Here and below, we assume that the laboratory frame moves
ultrarelativistically with respect to the comoving frame.}. Therefore, the observer will detect nearly all the
photons, apart from those emitted precisely toward
the star in the comoving frame.

\begin{figure}
\vspace*{-3cm}
\resizebox{0.5\hsize}{!}{\includegraphics{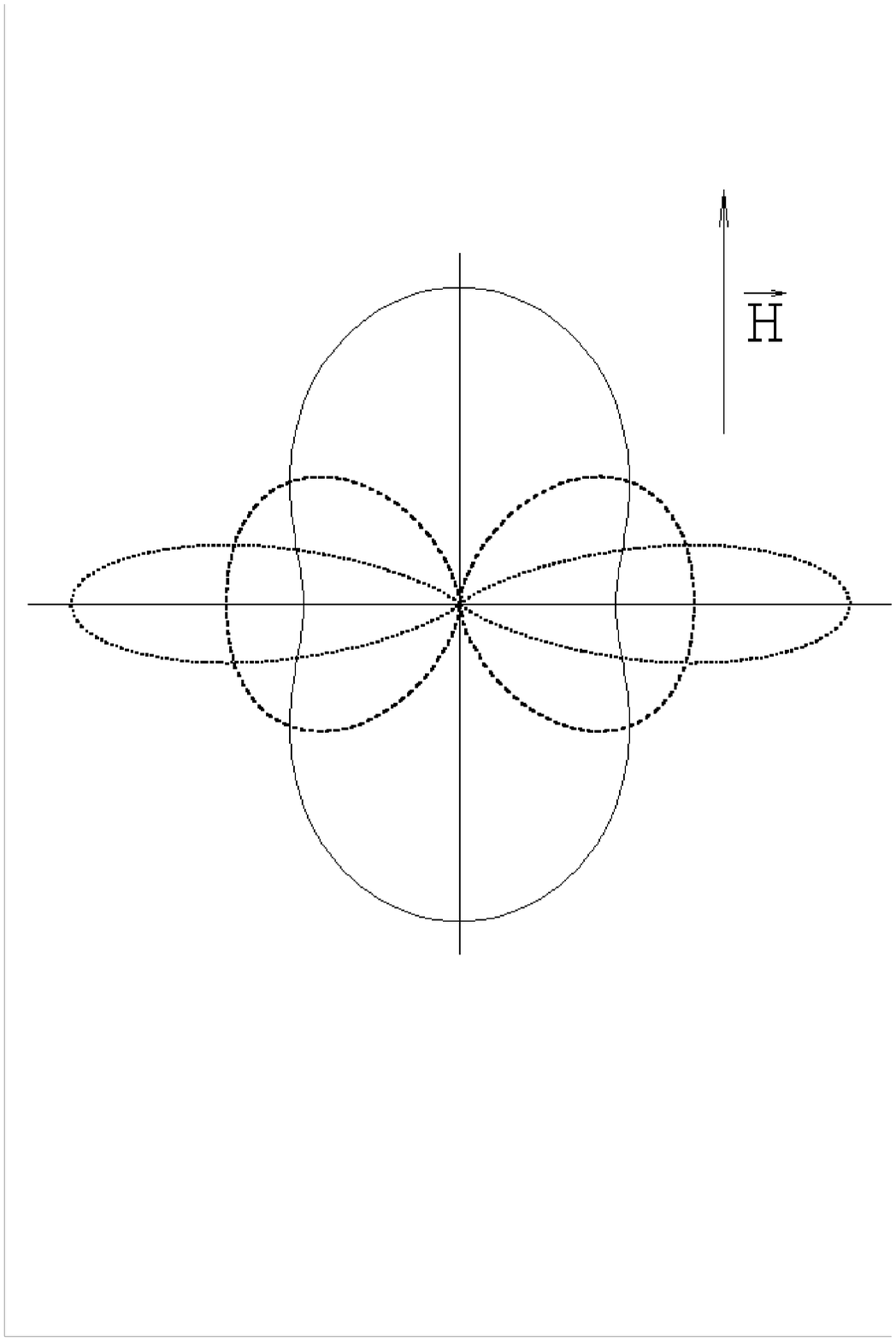}}
\vspace*{-3cm}
\caption{Diagram of the directivity of the first (solid curve),
second (dashed curve), and tenth (dotted curve) cyclotron
harmonics. The scales are arbitrary. The direction of the
magnetic field is marked by the arrow.}
\label{fig3}
\end{figure}

As was noted above, the energy of all photons corresponding to
a single harmonic is the same in the comoving frame.
Nevertheless, the coefficient of transformation
of the photon frequency from the comoving
to the laboratory frame is not constant, and depends
strongly on the angle $\theta'$. Using (\ref{d10}),
we can easily show that the frequency of photons emitted precisely
toward the observer ($\theta'\simeq 0$) increases by the factor $2 \gamma$,
the frequency of photons emitted perpendicular
to the direction toward the observer
($\theta'\simeq \pi/2$ in the
comoving frame) increases by the factor $\gamma$, and the
frequency of photons emitted at the angle $\theta' = \pi-\frac{1}{\gamma}$
decreases by the factor $\gamma$. This obviously leads to
considerable broadening of the lines. In addition, the
equidistant character of the lines will be disrupted.
This can be interpreted as follows. According to (\ref{d1})),
most photons associated with the first harmonic are
emitted at angles $\theta'\simeq 0$.Therefore, their frequency
increases by the factor $2 \gamma$, and the maximum of the
first harmonic occurs at $2 \gamma \omega_H$. On the other hand,
most photons associated with the second and higher
harmonics are emitted perpendicular to the direction
toward the observer ($\theta'\simeq \pi/2$), and particles with
$\theta'\simeq 0$ are virtually absent (Fig.~\ref{fig3}).
The frequencies
of these photons increase only by the factor $\gamma$, so that
the maxima of higher harmonics are located at
\begin{equation}
\label{f7}
\omega = \gamma \omega_n = \gamma n \omega_H; n>1
\end{equation}
In particular, the maxima of the first and second harmonics
coincide. Since the harmonics overlap, their
maxima are shifted, and relation (\ref{f7}) is not precisely
satisfied. The resulting spectrum differs substantially
from a spectrum with equidistant features.

We have considered above only radiation by electrons
whose velocities along the field are exactly the
same. At the same time, a real ensemble of electrons
will certainly be characterized by some distribution of
the momentum component along the field. Therefore,
the total velocity distribution of the emitting electrons
can be written
\begin{equation}
\label{j6}
dn = f(p_\parallel) exp\left(-\frac{m_e \upsilon^2}{2 T}\right)\,
dp_\parallel d\left(-\frac{m_e \upsilon^2}{2 T}\right)
\end{equation}
This distribution will considerably affect the observed
spectrum of an X-ray pulsar. In particular, the first
harmonic may be transformed from a sharp peak (as
in Fig. 2) to a broad line. Therefore, it is interesting
to answer the following question: can the distribution $f(p_\parallel)$ in
(\ref{j6}) be very broad for a real source, and can
it result in the transformation of separate harmonics
into a continuous spectrum?

As is noted above, ultrarelativistic particles can be
produced by collisionless shocks formed in the accretion
flow. They do not leave the shock, and oscillate at
the shock front. Let us suppose that such oscillations
are roughly harmonic. Then, the particle momentum
will vary as $p = p_0 \cos\alpha t$, where $\alpha$ is the oscillation
frequency. Since the oscillations are ultrarelativistic,
the momenta of the particles are proportional to their
Lorentz factors $\gamma$ ($p\sim\gamma$),
so that the $\gamma$ values vary in the same way:
\begin{equation}
\label{w1}
\gamma = \gamma_0 \cos\alpha t
\end{equation}
The distribution of the corresponding ensemble of
particles along the field can be represented
\begin{equation}
\label{w5}
dn = \frac{N}{\pi \gamma_0 \sqrt{1-\left(1-\frac{\gamma}{\gamma_0}\right)^2}} d\gamma;\quad \gamma\in[1;\gamma_0]
\end{equation}
i.e., the distribution is characterized by a sharp peak
with its maximum at $\gamma\simeq\gamma_0$. Of course, real distributions
can differ considerably from (\ref{w5}); nevertheless,
they can be sufficiently narrow that the individual
cyclotron harmonics are not transformed into a continuous
spectrum.

\section{DISCUSSION}
The parameters of the four pulsars for which
higher cyclotron harmonics have been detected are
presented in the table. Let us consider in more detail
the high-energy spectra of these sources, since they
contain the features of the most interest to us. The
spectrum in this region possesses a quasi-power-law
character with an exponential cut-off (if cyclotron
features are not taken into consideration), as is
very typical for X-ray pulsars. Spectra of this form
are supposed to be produced by comptonization of
relatively cool radiation by hot electrons, considered
in detail \cite{y}. The resultingspec trum for energies
above the cut-off at $E_c$ is approximately a Wien
spectrum with the characteristic temperature equal
to the temperature of the hot electrons $T_e$ [15]. It
is formed by photons that have undergone a large
number of collisions with the electrons, and have
therefore been heated to the temperature $T_e$. On the
other hand, the photons forming the power-law part
of the spectrum have experienced only a few collisions
with hot electrons, and have not reached thermal
equilibrium. As a result, the power-law spectrum at
each specific frequency contains fewer photons than
a Planck spectrum with temperature $T_e$.

To form an absorption line in such a spectrum, the
electrons must have a temperature below $T_e$. Even
if we assume that the electron-velocity distribution
differs considerably from maxwellian, the situation is
unlikely to change fundamentally. In any case, the
characteristic energy of motion of the electrons across
the magnetic field should not exceed $T_e$. The table
shows that, for all four sources, the temperature $T_e$
does not exceed $20$~{keV}. On the other hand, the temperature
of the electrons forming complex cyclotron
lines must be considerably greater: for either emission
or absorption harmonics of higher orders to appear
in the spectrum, the velocity of the electrons across
the magnetic field must be comparable to the speed
of light; i.e., their temperature must be comparable to
$mc^2 \simeq 500$~{keV}. This requirement can be explained
quantitatively. The ratio of the optical depths of the
ultrarelativistic electrons to cyclotron emission (or
absorption) at two neighboring harmonics is of the
order of the ratio of the mean kinetic energy of motion
of the particles across the magnetic field to their rest
energy. We can easily demonstrate using(\ref{d1}) that, if
the motion of the electrons across the field is slow
 ($\upsilon \ll c$), this ratio is \footnote{7In fact, when at least one absorption line is observed in
the non-planckian part of the spectrum, formula (\ref{e14}) is not
completely accurate, so that the deviations of the source
spectrum from a Planck spectrum must be taken into account
when estimating the temperature of the absorbing
electrons from a ratio of relative line intensities. However,
for the four objects under consideration, our conclusion that
the cyclotron features in their spectra are emission features
seems to remain valid.}
\begin{equation}
\label{e14}
\frac{\tau_{n+1}}{\tau_n}=\frac{n (n+2)}{n+\frac{3}{2}}\left(1+\frac{1}{n}\right)^{2n+2}
\left(\frac{k T}{2 m_e c^2}\right)
\end{equation}
Consequently, if the temperature satisfies the condition
\begin{equation}
\label{e15}
T < T_e \sim 20 \mbox{keV}
\end{equation}
then the intensity of harmonics should decrease
rapidly with increasing harmonic number. In particular,
the optical-depth ratio for the second and
first harmonics should not exceed $\frac{2}{5}$. However, observed
values of this ratio (presented in the table)
are $2.7$ for X0115+63, $9.3$ for 4U1907+09, $> 1.8$
for Vela X-1, and $2.8$ for A0535+26 (the cyclotron
features were considered in all cases to be absorption
features). According to (\ref{e14}), this corresponds to
electron temperatures of $141$, $483$, $94$, and $146$~{keV},
respectively. It is unclear how these hot electrons can
absorb radiation with a temperature of $\sim 20$~{keV}.
Consequently, the cyclotron features observed in the
spectra of these objects must be emission lines.
\begin{figure}
\begin{tabular}{|l|c|c|c|c|}
\hline
\hline
Source&X0115+63&Vela X-1&4U1907+09&A0535+26\\
\hline
Reference&\cite{b}&\cite{c}&\cite{c2}&\cite{y1}\\
\hline
Temperature                               &    &   &  &    \\
            of the Planck                  &17,4&9,4&12&18,7\\
                         tail $T_e$(keV)&    &   &  &    \\
\hline
Ratio              &   &         &   &   \\
of optical depths       &2,7&1,8 - 9,8&9,3&2,8\\
of the first and second        &   &         &   &   \\
harmonics $\tau_2/\tau_1$&   &         &   &   \\
\hline
\end{tabular}
\end{figure}

As was reported in \cite{d1} and \cite{c} the cyclotron lines in the
spectra of X0115+63 and Vela X-1 are not equally
spaced (the lines were considered in both papers to
be absorption features). Let us discuss X0115+63 in
more detail. The observed lines in its spectrum \cite{b}
are not equidistant (if they are considered to be emission
features). In addition, the lines are very broad
and overlapping. All these characteristic features of
the cyclotron spectrum are usually explained by the
complex geometry of the emitting region (a hot spot
on the surface of the neutron star). The broadening
of the line and its transformation to a broad band
are thought to be due to variations in the intensity
of the magnetic field within the emitting region\footnote{As is noted above,
in the nonrelativistic case, cyclotron
radiation (absorption) occurs at a single frequency, which
depends only on the magnetic-field intensity, in accordance
with (\ref{f5}).}, while the higher harmonics are formed in parts of the
hot spot where the magnetic field is weaker than in
the regions of formation of the main harmonic, so
that their frequencies are shifted with respect to the
fundamental frequency, $\omega_n < n\omega_1$.

Unfortunately, this explanation runs into serious
problems. It is not difficult to estimate the relative
broadening of lines due to the above mechanism.
First, we have
\begin{equation}
\label{mistake}
\frac{\Delta\omega}{\omega} = \frac{\Delta H}{H}
\end{equation}
Let us assume that a neutron star is characterized
by the standard radius ($R = 10~{ª¬}$) and mass
($M = 1.4 M_\odot$), a dipolar magnetic field
\begin{equation}
\label{f6}
H_r = \frac{\bf m}{r^3} \cos \theta; \qquad
H_\theta = -\frac{\bf m}{2 r^3} \sin \theta;
\end{equation}
and a temperature at the foot of the accretion column $T_e\sim 10$~{keV}.
Then, the height of the hot region is approximately determined by the usual
barometric formula
$$
\Delta r = \frac{k T_e}{m_H g}
$$
where $k$ is the Boltzmann constant, $m_H$ the mass
of the hydrogen atom (the dominant component of
the accretion gas), and $g$ the free-fall acceleration at
the neutron-star surface. Substituting the numerical
values, we obtain
$\Delta r \sim 0.7~{¬}$. The relative variation
in the magnetic field of the form (\ref{f6})
within this distance from the neutron-star surface is
$$
\frac{\Delta H}{H} = 3 \frac{\Delta r}{R} \sim 2 \cdot 10^{-4}
$$
Consequently, $\displaystyle{\frac{\Delta\omega}{\omega} \sim 2 \cdot 10^{-4}}$.
Further, let us estimate the variations in the magnetic field over the spot.
If the angular scale of the spot is $\chi$, then, as obviously
follows from (\ref{f6}), the corresponding relative variation
in the magnetic field will be
$$
\frac{\Delta H}{H} =  \frac{3}{8} \chi^2
$$
According to \cite{t1}, the angular size of the spot is
$$
\chi \simeq \sin \chi = 7,455 \cdot 10^{-4} H^{-\frac{2}{7}} L^{\frac{1}{7}}
$$
where $L$ is the luminosity of the X-ray pulsar.
The luminosity of the pulsar X0115+63 is $L \simeq 10^{37}$~{erg/s} \cite{a},
while the magnetic field (derived from
the energy of the cyclotron line using the nonrelativistic
formula) would be at least $ 7 \cdot 10^{11}$~{Gs}.
Hence, $\chi \simeq 0,06$, i.e. $\displaystyle{\frac{\Delta\omega}{\omega} =
\frac{\Delta H}{H} \approx 1,4 \cdot 10^{-3}}$.

\begin{figure}[p]
\resizebox{0.9\hsize}{!}{\includegraphics{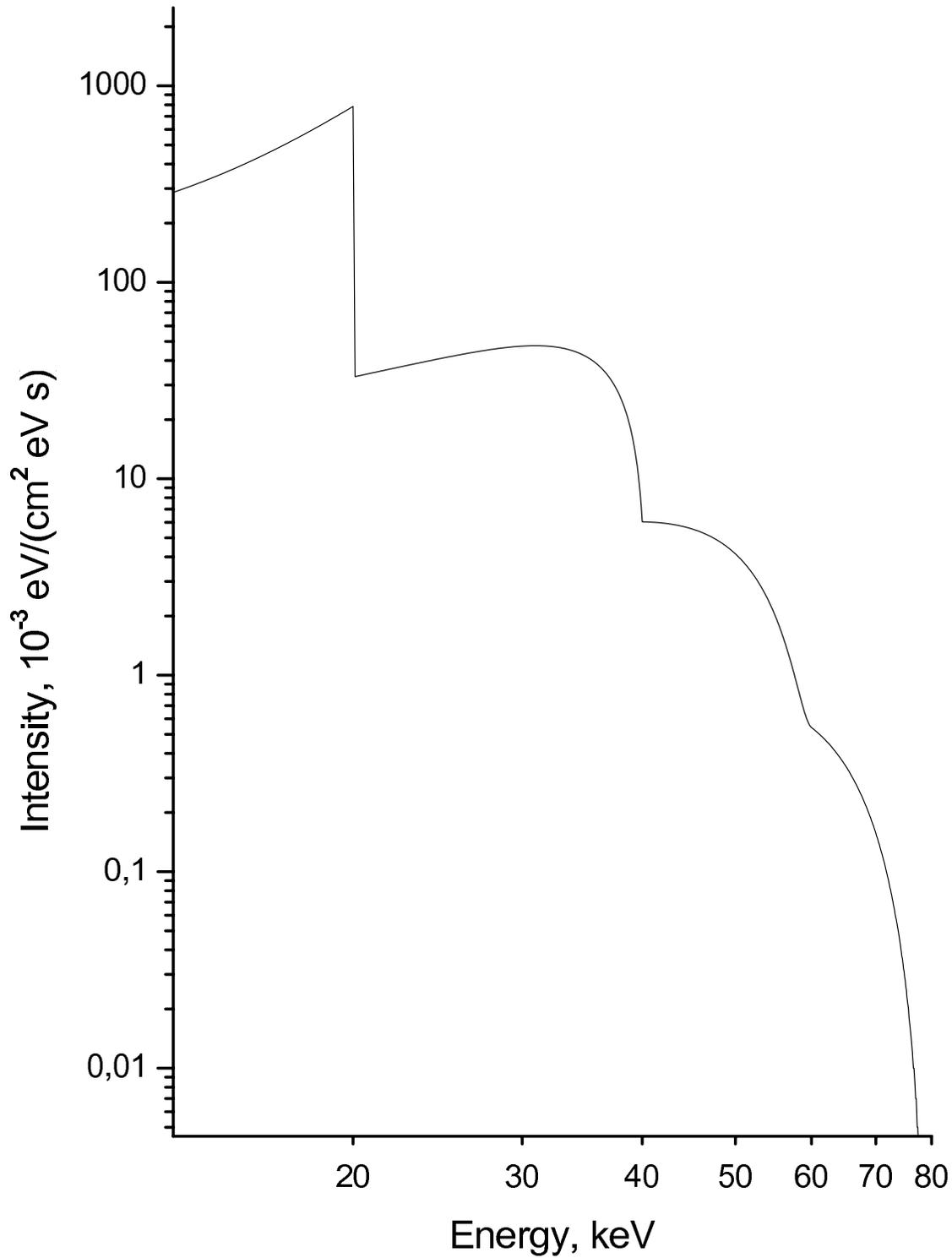}}
\caption{Energy spectrum of cyclotron radiation by an
X-ray pulsar containing four harmonics. The transverse
temperature of the emitting electrons is $T=20$~{keV}. The
scales are logarithmic along both axes.}
\label{fig4}
\end{figure}

Therefore, the possible broadening or shift of the cyclotron line
frequency due to the source geometry is
$\displaystyle{\frac{\Delta\omega}{\omega} \le 2 \cdot 10^{-3}}$.
Of course, this is completely inadequate to
explain the observed width of the lines
($\displaystyle{\frac{\Delta\omega}{\omega} \sim 0.2}$),
Moreover, to reproduce the non-equal spacing of the
lines, it is necessary to suppose a complex (and,
therefore, quite artificial) temperature distribution
over the spot. Thus, we cannot explain the observed width of the
cyclotron lines, their deviation from an equidistant
distribution, or their overlapping using only geometrical
considerations. The most probable and natural
explanation of all these properties is that the electrons
giving rise to the cyclotron lines are ultrarelativistic,
and so have a very anisotropic distribution of the form (\ref{j6}).
Radiation produced by such particles possesses
precisely these features. Therefore, the presence of
several broad cyclotron lines separated by unequal
distances in an X-ray pulsar's spectrum represents
a strong argument in favor of our mechanism for the
formation of the cyclotron features. The spectrum of
cyclotron radiation calculated using( (\ref{f1}).
is plotted in Fig.~\ref{fig4} on a log–log scale for a transverse temperature
of 20~{keV}. There is an obvious similarity between
the observed spectrum of X0115+63 in \cite{b}
and the calculated spectrum in Fig.~\ref{fig4}. Although the first harmonic
in Fig.~\ref{fig4} represents a sharp peak, whereas it
is quite broad in the observed spectrum, this could
be explained as an effect of the velocity distribution (\ref{j6})
of the emitting electrons along the magnetic field. This same distribution
should also result in additional broadening of the cyclotron lines.

The calculated cyclotron spectrum displays another
very interesting feature: as we can see in Fig.~\ref{fig2},
the dips between neighboring harmonics are located
at nearly equal distances from each other. Therefore,
if these decreases in intensity in the observed
spectrum were interpreted as absorption lines, they
would appear be equidistant. This is probably why
the cyclotron features in the spectra of X0115+63,
Vela X-1, 4U1907+09, and A0535+26 have been
interpreted as a set of equidistant absorption lines.
However, as is demonstrated above, they are more
likely to be emission rather than absorption lines.

We have already noted that ultrarelativistic, anisotropic
electrons can be produced by a collisionless
shock in the accretion flow. As is shown in \cite{n,t}
such shocks can indeed form, and the corresponding electrons
will have a strongly anisotropic velocity
distribution. This is due to the fact that the electrons
are accelerated primarily along the magnetic field. In
addition, the transverse component of their momenta
rapidly decreases due to radiative cooling, while the
component along the field changes relatively little.
The estimation of the electron temperature performed
in \cite{n,t} yielded values $T\sim 10$~{MeV}, corresponding
to $\gamma \sim 20$, i.e. the electrons are ultrarelativistic.

Note also that there is considerable indirect evidence
that the particles forming the cyclotron lines
in the spectra of X-ray pulsars are ultrarelativistic,
primarily associated with discrepancies in estimates
of the magnetic fields of these objects \cite{i}.

The pulsar Her X-1 was considered in detail in \cite{t1}. In general,
its rotation speeds up, but it experiences
deceleration in some intervals \cite{s}. Consequently,
its period is close to the equilibrium period;
i.e., the angular velocity of rotation of the neutron star
together with its magnetosphere is close to keplerian
at the Alfven radius. We can use this fact to estimate
the magnetic field of this pulsar, which turns out to
be relatively small, about $H = 3\cdot 10^{11}$~{Gs}. In addition,
radio pulsars in binary systems possess anomalously
weak magnetic fields, $H \simeq 10^{8} - 10^{11}$~{Gs} \cite{t2}. Their
formation as the result of the evolution of a pair containing an
X-ray pulsar could be naturally explained
if these pulsars had comparatively small magnetic
fields, $H < 3 \cdot 10^{11}$~{Gs}. Mihara et al. \cite{q}
explained the
observed 35-day cycle of Her X-1 as the result of periodic
eclipses of the emitting region on the neutron star
surface by the accretion disk. For this mechanism
to operate, the disk must be located at a sufficiently
short distance from the surface; i.e., the field must be $H < 10^{11}$~{Gs}.

At the same time, a cyclotron line at energy
$35–56$~{keV} is observed in the spectrum of Her X-
1. If the nonrelativistic formula (\ref{f5}), is applied, this
corresponds to a magnetic field of $(3-5) \cdot 10^{12}$~{Gs}.
This contradiction appears to be due to the fact that
the electrons emitting the cyclotron line are actually
ultrarelativistic. In this case, as is shown above, the
energy of the first (fundamental) harmonic increases
by the factor $2 \gamma$. Therefore, the nonrelativistic formula
gives a substantially (by a factor of $2 \gamma$) overestimated
value for the magnetic field. Thus, adopting the
hypothesis that the velocities of the electrons emitting
the cyclotron line are ultrarelativistic along the field
makes it possible to avoid contradictions between
estimates of the pulsar's magnetic field given by
various methods; this represents weighty indirect
evidence in favor of our mechanism for the formation
of the cyclotron radiation.

Another argument supporting this mechanism is
the observation of correlated variations in the energies
of the cyclotron lines of some pulsars and in their
luminosities \cite{z}. This correlation can be explain in
a natural way by our model. The luminosity of the
hot spot is comparable to the Eddington luminosity;
i.e., the radiation pressure appreciably affects the accretion
rate. If the luminosity increases, the accretion
flow will be substantially slowed. As a result, the
intensity of the shock decreases, and the mean energy
of the ultrarelativistic particles produced by the shock
decreases. Consequently, the energy of the cyclotron
lines also decreases \cite{i}.
However, we emphasize that
it is recent observations of higher harmonics and their
deviations from an equidistant spectral distribution
that represent the first direct evidence that the electrons
emitting the cyclotron lines are ultrarelativistic,
and have very anisotropic distributions of the form (\ref{j6}).

In this mechanism for the formation of the cyclotron lines, we must know the electron Lorentz
factor $\gamma$ to determine the pulsar's magnetic field. In turn, this factor depends on the physical conditions
in the accretion flow, in particular, in the collisionless shocks. It is possible that different types of X-ray pulsars
have different shock structures, and, therefore, different characteristic $\gamma$ values.

Accretion in low-mass binary systems occurs via the flow of material through the inner Lagrange point
when the donor star fills its Roche lobe, whereas accretion in massive systems occurs via the capture
of material from the powerful stellar wind of the O–B companion, which does not fill its Roche lobe. It
may not be a chance coincidence that all four sources displaying several cyclotron harmonics are associated
with massive systems, so that the second type of accretion is realized. In this case, the conditions at the
front of the collisionless shock may be more favorable for the ultrarelativistic electrons to acquire greater
momentum transverse to the magnetic field, resulting in the appearance of higher cyclotron harmonics in
the pulsar spectrum.

\section{CONCLUSION}
The presence of several cyclotron harmonics in the spectra of X-ray pulsars provides important information
about their physical properties.

First, the cyclotron features should apparently not be interpreted as absorption lines. The ratio of the
optical depths of harmonics of different orders can be used to estimate the temperature of the electrons participating in the formation
of the cyclotron features. For absorption to take place, this temperature
must be less than the temperature of the Planck tail that is often observed in the spectra of X-ray pulsars.
As a rule, $T_e$ is not large: for many pulsars, $T_e < 20~{KeV}$. As a result, the second harmonic is substantially
weaker than the first, and is difficult to detect observationally. In any case, the cyclotron features in
all four sources for which the second harmonic has been observed are very likely emission features.

Second, if the cyclotron emission lines are not equidistant, this argues strongly that the electrons
emitting these lines are ultrarelativistic and have a very anisotropic distribution. Emission by such electrons
displays characteristic properties that are not typical of emission by nonrelativistic particles, and
which are actually observed in the observed spectra of X-ray pulsars. There is also other (indirect) evidence
supporting this mechanism for the cyclotron-line formation. Thus, the cyclotron radiation of X-ray pulsars is
most likely produced by anisotropic ultrarelativistic electrons. As a result, the nonrelativistic formula
(\ref{f5}) considerably overestimates the pulsar magnetic fields. The magnitude of this overestimation depends
on the distribution function (\ref{j6}) of the emitting electrons in the corresponding shocks. We cannot rule out
the possibility that this function may be different for different types of X-ray pulsars. Subsequent studies
of the physical processes occurring in the accretion flows of these objects may help shed light on this problem.

Thus, investigations of the spectra of X-ray pulsars containing several cyclotron harmonics may enable us to refine estimates of their
magnetic fields, and thereby to resolve questions concerning the structure and evolution of these objects.

\section{ACKNOWLEDGEMENTS}
I am grateful to G.S. Bisnovatyi-Kogan for help
in preparation of this article and, in particular, for
several fruitful ideas. This work was supported by
the Russian Foundation for Basic Research (project
codes 01-02-06146 and 99-02-18180).

\end{document}